\documentclass[11pt]{article}
\usepackage{hyperref}
\usepackage{graphicx}
\usepackage{graphics}
\usepackage{dsfont}
\usepackage{epsfig}
\usepackage{amsmath,amssymb,amsthm,amscd}
\allowdisplaybreaks

\newcommand{\tr}{\textrm{tr}}

\numberwithin{equation}{section}

\begin{document}
\begin{titlepage}
{}~ \hfill\vbox{ \hbox{} }\break

\rightline{USTC-ICTS-13-08}

\vskip 3 cm

\centerline{\Large 
\bf  
Dijkgraaf-Vafa conjecture and $\beta$-deformed matrix models}   
\vskip 0.5 cm

\renewcommand{\thefootnote}{\fnsymbol{footnote}}
\vskip 30pt \centerline{ {\large \rm Min-xin Huang
\footnote{minxin@ustc.edu.cn}  } } \vskip .5cm \vskip 30pt

\begin{center}
{Interdisciplinary Center for Theoretical Study,  \\ \vskip 0.2 cm  University of Science and Technology of China,  Hefei, Anhui 230026, China} 
\end{center}

\setcounter{footnote}{0}
\renewcommand{\thefootnote}{\arabic{footnote}}
\vskip 60pt
\begin{abstract}

We study the  $\beta$-deformed matrix models using the method of refined topological string theory. The refined holomorphic anomaly equation and boundary conditions near the singular divisors of the underlying geometry fix the refined amplitudes recursively. We provide exact test of the quantum integrality conjecture in the Nekrasov-Shatashvili limit. We check the higher genus exact formulae with perturbative matrix model calculations.

\end{abstract}

\end{titlepage}
\vfill \eject


\newpage

\baselineskip=16pt

\tableofcontents

\section{Introduction and Summary}

Recently there have been some interests in refined topological string theory, which originate from the $\Omega$ deformation in supersymmetric gauge theory \cite{NO}. Certain matrix models for the $\mathcal{N}=2$ gauge theory and refined topological string theory are derived and studied in e.g. \cite{DV:2009, Sulkowski:2009, Sulkowski:2010}.  It is expected that the Dijkgraaf-Vafa conjecture \cite{DV2002}, which relates matrix models with topological strings on certain local Calabi-Yau manifolds, can be generalized to the refined case. Here the refinement will correspond to the $\beta$-deformation of the matrix models. The topological expansion of ordinary matrix model free energy has been constructed from the spectral curves \cite{Akemann:1996zr, EO}, and can be generalized to the $\beta$-deformed case \cite{Chekhov:2006, Chekhov:2010}. However, the topological recursion method is still too difficult for many practical calculations in the $\beta$-deformed case, see e.g. \cite{Brini:2010, Krefl:2012}. Furthermore, the holomorphic anomaly equation can be derived from the topological recursion in the undeformed case \cite{Eynard:2007hf}. For the $\beta$-deformed case such a derivation is not available at the moment in the literature. In this paper,  we shall study the simple example of $\beta$-deformed cubic matrix model. Continuing in the direction of previous works \cite{HK:2006, Klemm:2010, Aganagic:2011}, we provide some higher genus formulae from the refined topological string method of holomorphic anomaly equation and the gap boundary conditions near singular divisors \cite{BCOV, Krefl:2010, Huang:2010}.

\section{Perturbative calculations}
The partition function $Z$ and free energy $F$ of a Hermitian matrix model with polynomial potential $W(\Phi)$ is defined by
\begin{eqnarray}
Z= e^{F} = \frac{1}{\textrm{Vol}(U(N)) }\int  D\Phi e^{-\frac{\tr W(\Phi)}{g_s}} = \frac{1}{N! (2\pi)^N }\int (\prod_{i} d\lambda_i )\Delta^2(\lambda) e^{-\sum_i \frac{W(\lambda_i )}{g_s} }, 
\end{eqnarray}
where $g_s$ is a perturbative expansion parameter, $\lambda_i ~ (i=1,2\cdots, N)$ are the eigenvalues  of the Hermitian matrix $\Phi$, and $\Delta(\lambda)=\prod_{i<j} (\lambda_i-\lambda_j)$ is the well known Vandermonde determinant. The $\beta$-deformation replaces the matrix integrand by its $\beta$ power, so the partition function becomes
\begin{eqnarray}
Z (\beta) = e^{F(\beta) } = \frac{1}{N! (2\pi)^N }\int (\prod_{i} d\lambda_i )\Delta^{2\beta} (\lambda) e^{-\frac{\beta}{g_s} \sum_i W(\lambda_i )}. 
\end{eqnarray}

In this paper we mostly consider a cubic polynomial potential
\begin{eqnarray} \label{potential}
 W(\Phi) =\frac{1}{2} \Phi^2 + \frac{1}{3} \Phi^3 .
 \end{eqnarray}
 We can compute the free energy perturbatively for small $g_s$, by expanding the exponential and reducing the computation to the expectation values of multi-trace operators in Gaussian matrix model. For the correspondence with Dijkgraaf-Vafa conjecture and topological string theory, we consider two-cut solution of the matrix model in the large $N$ limit. The eigenvalues of the matrix $\Phi$ distribute continuously around the two extrema of the potential (\ref{potential}). Here one of the extrema is actually a local maximum, so we need to use an analytic continuation to  anti-Hermitian matrix, so that the expansion around the  local maximum is the usual Gaussian matrix model. 

The perturbative calculations in the case without $\beta$-deformation, i.e. $\beta=1$, were studied in detail in \cite{KMT}. The idea is similar in the $\beta$-deformed case, and the partition function of the two-cut solution can be written as a two-matrix model  
\begin{eqnarray} \label{twomatrix}
Z =  \frac{1}{N_1! (2\pi)^{N_1} N_2! (2\pi)^{N_2}} \int (\prod_{i=1}^{N_1} d\mu_i )\Delta^{2\beta} (\mu) 
(\prod_{i=1}^{N_2} d\nu_i )\Delta^{2\beta} (\nu)  e^{-W_1(\Phi_1) - W_2(\Phi_2) - W(\Phi_1,\Phi_2)},
\end{eqnarray}
where $\mu_i$ and $\nu_j$ are the eigenvalues of the two Hermitian matrices $\Phi_1, \Phi_2$, and the potentials  are 
\begin{eqnarray} \label{potentials}
W_1(\Phi_1) &=& \tr [ \frac{1}{2} \Phi_1^2 +\frac{1}{3} (\frac{g_s}{\beta})^{\frac{1}{2}} \Phi_1^3 ],   \nonumber \\
W_2(\Phi_2) &=& - \tr [\frac{1}{2} \Phi_2^2 -\frac{1}{3}   (\frac{g_s}{\beta})^{\frac{1}{2}} \Phi_2^3],   \nonumber \\
W(\Phi_1,\Phi_2) &=& \frac{\beta N_2}{6 g_s}  - \beta N_1N_2 \log(\frac{g_s}{\beta})  \nonumber \\ && 
+2\beta \sum_{k=1}^{\infty} \frac{1}{k}  (\frac{g_s}{\beta})^{\frac{k}{2}} \sum_{p=0}^{k} (-1)^p \binom{k}{p} \tr(\Phi_1^p ) \tr(\Phi_2^{k-p} ).
\end{eqnarray}
Here  the interaction term  comes  from the exponentiation of the interaction of the eigenvalues $\mu_i$ and $\nu_j$ in the Vandermonde determinant. We have rescaled the matrices $\Phi_1$ and $\Phi_2$ by a factor $ (\frac{g_s}{\beta})^{\frac{1}{2}}$ to normalize the Gaussian potential, and neglected the unimportant overall factor in the partition function from the scaling.  

The correlators of multi-trace operators can be computed recursively by the loop equation of the $\beta$-deformed matrix model \cite{Mironov:2011, Morozov:2010}. Denote the correlators 
\begin{eqnarray}
C_{k_1,k_2, \cdots, k_m}(N,\beta) = \frac{\int D_\beta \Phi \tr(\Phi^{k_1}) \tr(\Phi^{k_2}) \cdots \tr(\Phi^{k_m}) \exp(-\frac{\tr\Phi^2}{2}) }{\int D_\beta \Phi \exp(-\frac{\tr\Phi^2}{2}) }, 
\end{eqnarray}
where $\Phi$ is a $N\times N$ Hermitian matrix, and  the $\beta$-deformed measure is $D_\beta \Phi\sim (\prod_{i} d\lambda_i )\Delta^{2\beta} (\lambda)$. Here the indices $k_i$ are non-negative integers and the correlator is understood to be zero by convention if one of the indices is negative.  For the cases of odd $\sum_{i=1}^m k_i $, the correlators vanish due to the reflection symmetry of the Gaussian potential. Also obvious is the situation of one of the indices  $k_i=0$, where we can simply pull out a  $N$ factor since $\tr(1)=N$. The loop equation, which is the Ward identity of matrix model, provides the following recursion relation for the correlators 
\begin{eqnarray} \label{recursion}
C_{k_1,k_2,\cdots, k_m} &=& \beta \sum_{i=0}^{k_m-2} C_{k_1,\cdots, k_{m-1}, i, k_m-2-i} +
\sum_{i=1}^{m-1} k_i C_{k_1,\cdots, k_i+k_m-2,\cdots,k_{m-1}} \nonumber 
\\ && + (1-\beta) (k_m-1) C_{k_1,\cdots, k_{m-1},k_m-2} . 
\end{eqnarray} 
Here the index $k_m$ for the recursion should be strictly positive $k_m>0$. For the undeformed case $\beta=1$, the above recursion can be easily understood from the Feynman rule of $m$ vertices, where a line of the last vertex can be contracted with another line from itself or a line from another vertex.  From the recursion relation (\ref{recursion}) and the initial conditions $C_{0} =N$, one can then compute the correlators recursively \cite{Morozov:2010}.  
 
We would like to obtain the genus expansion of the free energy. It turns out that there is a nice shift of the t'Hooft parameters so that the odd terms in the expansion vanish \cite{Aganagic:2011}. This is similar to the shift of mass parameters $m_i\rightarrow m_i +\frac{\epsilon_1+\epsilon_2}{2}$ in Seiberg-Witten theory studied in \cite{Krefl:2010jb, Huang:2011}. 

There are two contributions to the free energy, known as the non-perturbative part and the perturbative part. The non-perturbative contributions come from the measure of the matrix integral and the $U(N)$ volume factor \cite{OV}, which is obtained by evaluating the partition function (\ref{twomatrix}), without the interaction term $W(\Phi_1,\Phi_2)$ and setting $g_s=0$. The case of $\beta$ deformation is computed in \cite{Brini:2010}. Up to an unimportant constant from the analytic continuation to Gaussian potential $\Phi_2\rightarrow i\Phi_2$, the result is
\begin{eqnarray} \label{Fnp}
F_{n.p.} &=& \sum_{i=1}^2 [\frac{\beta^2 t^2_i}{2} (\log(t_i)-\frac{3}{2}) g_s^{-2} +\frac{\beta-1}{2} (\log(\beta t_i) -1) t_i g_s^{-1} 
\nonumber \\   &&  +\frac{1-3\beta+\beta^2}{12\beta} \log(\beta t_i) 
+\frac{1-\beta}{24\beta t_i} g_s + \frac{1-5\beta^2 +\beta^4}{720 \beta^3 t_i^2} g_s^2 + \mathcal{O}(g_s^3) ], 
\end{eqnarray}
where the t'Hooft parameters are defined as 
\begin{eqnarray}
t_i= g_s N_i,~~~~  i=1,2 . 
\end{eqnarray}

We see naively there are odd $g_s$ power terms in (\ref{Fnp}). In refined topological string theory, the free energy comes from the effective action of integrating out charged BPS particles in graviphoton background, and as such only even power terms appear. The situation is remedied by a shift 
\begin{eqnarray} \label{shift}
t_i\rightarrow t_i - \frac{\beta-1}{2\beta} g_s, ~~~ i=1,2 . 
\end{eqnarray}
The shift cancels the odd power terms in the non-perturbative contribution (\ref{Fnp}). We use the calligraphic symbol to denote the free energy after the shift 
\begin{eqnarray}
\mathcal{F}_{n.p.} = \sum_{i=1}^2 [ (\frac{\log t_i}{2} -\frac{3}{4}) \frac{\beta t_i^2}{g_s^2}  - \frac{\beta^2+1}{24\beta} \log(t_i) 
-\frac{7\beta^4+10\beta^2+7 }{5760 \beta^3} \frac{g_s^2}{t_i^2} +  \mathcal{O}(g_s^4) ]. 
\end{eqnarray}  
 
Next we compute the perturbative contributions.  We can neglect the unimportant terms in the first line in $W(\Phi_1,\Phi_2)$ in (\ref{potentials}), and keep the  contributions that are  perturbative for small $g_s$. We expand the remaining exponents in (\ref{twomatrix}), and use the recursion (\ref{recursion}). We find the first few terms 
\begin{eqnarray}
F_{pert}  &=&  (N_1-N_2) [ 2  \beta^2 (2 N_1^2 - 13 N_1 N_2 + 2 N_2^2)  - 
 9  \beta (\beta-1) (N_1 + N_2) 
 + (5\beta^2  - 9 \beta +  5 )  ] \frac{g_s}{6 \beta} \nonumber \\ &&  + [2\beta^3  (8 N_1^4 - 91 N_1^3 N_2 + 177 N_1^2 N_2^2 - 91 N_1 N_2^3 +   8 N_2^4)   
 -   \beta^2 (  \beta-1 ) (59 N_1^3 - 81 N_1^2 N_2  \nonumber \\ && - 81 N_1 N_2^2    + 
    59 N_2^3)  + \beta  (73 \beta^2 - 132 \beta + 73 ) (N_1^2 +   N_2^2) 
     -  2 \beta (92 \beta^2 - 153 \beta + 92)N_1 N_2   \nonumber \\ && 
       - ( \beta-1) (30 \beta^2 - 43 \beta + 30) (N_1 + N_2)  ]\frac{g_s^2}{6 \beta^2}  +\mathcal{O}({g_s^3}). 
\end{eqnarray} 
 
Again we find that the shift (\ref{shift}) eliminates the odd $g_s$ power terms in the perturbative free energy. Putting the non-perturbative and perturbative contributions (after the shift) together, we can extract the higher genus refined amplitudes $\mathcal{F}^{(n,g)}$ by the expansion 
\begin{eqnarray}
\mathcal{F}_{n.p.} + \mathcal{F}_{pert} = \sum_{n,g=0}^{\infty}  g_s^{2(n+g)-2} \beta^{1-g-2n} (\beta-1)^{2n} \mathcal{F}^{(n,g)}(t_1,t_2) . 
\end{eqnarray}
  
We list the first few refined topological amplitudes. Here the unrefined cases of $F^{(0,g)}$ have been computed in \cite{KMT}, and for completeness we also display them here. 
\begin{eqnarray}   \label{perturbativeF00}
\mathcal{F}^{(0,0)} &=& \sum_{i=1}^2 (\frac{t^2_i \log t_i}{2} -\frac{3t_i^2}{4}) + ( \frac{2}{3} t_ 1^3- 5 t_ 1^2 t_ 2 + 5 t_ 1 t_ 2^2 -  \frac{2}{3} t_ 2^3) + 
 \frac{1}{3} (8 t_ 1^4 - 91 t_ 1^3 t_ 2  \nonumber \\  &&
 + 177 t_ 1^2 t_ 2^2 -  91 t_ 1 t_ 2^3 + 8 t_ 2^4) +\mathcal{O}(t^5) , \\ \label{perturbativeF10}
 \mathcal{F}^{(1,0)} &=& - \sum_{i=1}^2 \frac{ \log t_i}{24}  
+ \frac{19}{12} (t_ 1 - t_ 2)  +  \frac{1}{6} (97 t_ 1^2 - 265 t_ 1 t_ 2 + 97 t_ 2^2)  
\nonumber \\  &&
+ \frac{1}{18} (4004 t_ 1^3 - 19401 t_ 1^2 t_ 2 + 19401 t_ 1 t_ 2^2 - 
    4004 t_ 2^3) +\mathcal{O}(t^4) ,
 \\     \label{perturbativeF01}
 \mathcal{F}^{(0,1)} &=& - \sum_{i=1}^2 \frac{ \log t_i}{12}  +\frac{1}{6} (t_ 1 - t_ 2)  + 
 \frac{1}{3} (7 t_ 1^2 - 31 t_ 1 t_ 2 + 7 t_ 2^2)  \nonumber \\  &&  + 
 \frac{1}{9} (332 t_ 1^3 - 2769 t_ 1^2 t_ 2 + 2769 t_ 1 t_ 2^2 -  332 t_ 2^3)      +\mathcal{O}(t^4)  \\  
 \mathcal{F}^{(2,0)} &=& - \sum_{i=1}^2 \frac{7}{5760 t_i^2}  +
 \frac{131}{48} + \frac{22709}{144} (t_ 1 - t_ 2)  + 
 \frac{1}{96} (581203 t_ 1^2 - 1449550 t_ 1 t_ 2 + 581203 t_ 2^2)  \nonumber \\  && 
 +\frac{23420099 (t_ 1^3+t_2^3) - 100452495 t_ 1 t_ 2 (t_1-t_2) }{120}    +\mathcal{O}(t^4)    \label{perturbativeF02} \\ 
 \mathcal{F}^{(1,1)} &=& - \sum_{i=1}^2 \frac{7}{1440 t_i^2}  + \frac{17}{12} +  \frac{4133}{36} (t_ 1 - t_ 2)  + 
 \frac{1}{24} (120367 t_ 1^2 - 342334 t_ 1 t_ 2 + 120367 t_ 2^2)   \nonumber \\  &&  + 
 \frac{5206799 (t_ 1^3 +t_2^3) - 26406615 t_ 1 t_ 2(t_1-t_2)  }{30}   +\mathcal{O}(t^4)     \label{perturbativeF11} \\   
 \mathcal{F}^{(0,2)} &=& - \sum_{i=1}^2 \frac{1}{240 t_i^2}    + \frac{35}{6} (t_ 1 - t_ 2) t + (338 t_ 1^2 - 1632 t_ 1 t_ 2 + 
    338 t_ 2^2)   \nonumber \\  &&  + 
 \frac{4}{5} [16533 (t_ 1^3 +t_2^3) - 151100 t_ 1 t_ 2(t_1-t_2)  ]  +\mathcal{O}(t^4).     \label{perturbativeF20}
\end{eqnarray}
The leading terms in $\mathcal{F}^{(n,g)}$ from the non-perturbative contributions are basically the singular terms appearing in integrating out a massless BPS particles in the general graviphoton background near a conifold point  in refined topological string theory. We note that there is a $(-1)^n$ factor difference  comparing with the convention in \cite{Huang:2010}, probably due to some analytic continuation between these two kinds of calculations.  
  
\section{Review of Dijkgraaf-Vafa geometry}

Dijkgraaf and Vafa conjectured \cite{DV2002} that the Hermitian matrix model with a degree $n+1$ polynomial potential $W(\Phi)$ is dual to the topological string theory on a local Calabi-Yau three-fold geometry, defined by the following curve on the $\mathbb{C}^4$ coordinates $(u,v,y,x)$  
\begin{eqnarray}
uv = y^2 + W^\prime(x)^2 + f(x), 
\end{eqnarray}
where $f(x)$ is a degree $n-1$ polynomial whose coefficients parametrize the complex structure moduli of the Calabi-Yau geometry.  The A-periods of the Calabi-Yau geometry are identified with the filling fractions, i.e. the t'Hooft parameters in large $N$ limit,  around the extrema of the potential in the $n$-cut  solution of the matrix model.  

We specialize to the case of cubic potential $W(x)$ in (\ref{potential}) and introduce some notations. The complex deformation $f(x)$ split the double roots $(a_1,a_2)$ in the equation $W^\prime(x)^2 + f(x) =0 $, whose roots are denoted as $(a_1^-, a_1^+, a_2^-,a_2^+) \equiv  (x_1,x_2,x_3,x_4)$. We define the complex parameters 
\begin{eqnarray}
z_1 = \frac{(x_2-x_1)^2}{4}, ~~~  z_2 = \frac{(x_4-x_3)^2}{4}, 
\end{eqnarray} 
which parametrize the complex structure moduli space of the Calabi-Yau geometry. The moduli space and its singular divisors are studied in detail in \cite{HK:2006}. Here the divisor $z_1z_2=0$ behaves like the conifold divisor. There are two other singular divisors 
\begin{eqnarray} \label{singulardivisors}
I (z_1,z_2) &\equiv& \frac{1}{4} [(x_3+x_4)-(x_1+x_2)]^2 = 1-2(z_1+z_2), \nonumber \\ 
J(z_1,z_2)  &\equiv& (x_1-x_3)(x_2-x_3)(x_1-x_4)(x_2-x_4)  \nonumber \\ 
&=& 1-6(z_1+z_2) +9z_1^2+14z_1z_2 +9z_2^2,
\end{eqnarray} 
where the singular divisor $J=0$ also behave like the conifold divisor, and the higher genus topological string amplitudes expended around the divisor will have gap like behavior \cite{Haghighat:2008, Klemm:2010}.  On the other hand, the singular divisor $I=0$ is actually an essential singularity where the perturbative solutions to the Picard-Fuchs equation have zero radius of convergence. 

The planar free energy $\mathcal{F}^{(0)}$ of the matrix model are determined by the well known equation $\frac{\partial \mathcal{F}^{(0)}}{\partial t_i} = \Pi_i$, where the periods are 
\begin{eqnarray} \label{periods3.4}
t_i =\frac{1}{2\pi i} \int_{a_i^-}^{a_i^+} \lambda dx, ~~~~  \Pi_i =\frac{1}{2\pi i} \int_{a_i^+}^{\Lambda } \lambda dx, 
\end{eqnarray}  
as contour integrals over cycles of $x$-plane, with  the one-form differential $\lambda dx = \sqrt{W^\prime(x)^2 + f(x)} dx$. The expansion for the A-periods to the first few orders are 
\begin{eqnarray} \label{periods.6.24}
t_1 &=&  \frac{z_1}{4} - \frac{z_1}{8}  (2 z_1 + 3 z_2)  - 
 \frac{z_1}{32}  (4 z_1^2 + 13 z_1 z_2 + 9 z_2^2)+\mathcal{O}(z^4) ,  \nonumber \\
t_2 &=&  -\frac{z_2}{4} + \frac{z_2}{8}  (2 z_2 + 3 z_1)  +
 \frac{z_2}{32}  (4 z_2^2 + 13 z_1 z_2 + 9 z_1^2)+\mathcal{O}(z^4) . 
\end{eqnarray}

To compute the period integrals, one finds differential operators $\mathcal{L}(z_1,z_2)$ such that $\mathcal{L} \lambda$ is a total derivative of $x$ variable. One also needs to be careful about possible residua, because of which an operator $\mathcal{L}$ may not annihilate the periods even though $\mathcal{L} \lambda$ is a total derivative \cite{HK:2006}. More precisely, up to the second order differentials, there are three linearly independent operators whose actions on $\lambda$ are total derivatives, which are the followings  
\begin{eqnarray} \label{PFoperators}
\mathcal{L}_1 &=&  [ (3 - 2z_1 - 6z_2)\partial_{z_1}  - 2z_1(1- 2z_1 - 6z_2 )\partial_{z_1}^2  +(z_1\leftrightarrow z_2)] \nonumber \\ && 
 +2 (1-5 z_1 - 5 z_2+6z_1^2 +4z_1z_2+6z_2^2 )\partial_{z_1} \partial_{z_2} ,  \nonumber \\
 \mathcal{L}_2 &=&
 -6 + 36 (z_1 + z_2) - 6 (9 z_1^2 + 14 z_1 z_2 + 9 z_2^2) 
 +2(z_1+z_2)[-1 + 6 z_1 + 6 z_2  
 -11 z_1^2 - 10 z_1 z_2  \nonumber \\ && - 11 z_2^2+ 6  (z_1 - z_2)^2 (z_1 + z_2)   ] \partial_{z_1} \partial_{z_2}
 +[ (  7 z_1 - 3 z_2 -39 z_1^2 - 18 z_1 z_2 + 9 z_2^2 + 
  46 z_1^3  \nonumber \\ && + 62 z_1^2 z_2 + 26 z_1 z_2^2 - 6 z_2^3) \partial_{z_1} + 2z_1( 1 - 6 z_1 -6 z_2 + 13 z_1^2 + 14 z_1 z_2 + 5 z_2^2 -10 z_1^3  \nonumber \\ && - 6 z_1^2 z_2 + 10 z_1 z_2^2 + 6 z_2^3) \partial_{z_1}^2 +(z_1\leftrightarrow z_2) ],  \nonumber \\ 
   \mathcal{L}_3 &=& [ (5z_1 + 3z_2) \partial_{z_1} + 2 z_1 (2 - 5z_1 - 3z_2) \partial_{z_1}^2 - (z_1\leftrightarrow z_2) ]
\nonumber \\ &&   - 2 (z_1 - z_2) (1 - 3  z_1 - 3 z_2) \partial_{z_1} \partial_{z_2}
 \end{eqnarray} 
All three operators annihilate the A-periods $t_i$. After adding proper constants from regularizing the integrals, the B-periods $ \Pi_i$ can be annihilated by the operators $\mathcal{L}_1$ and $\mathcal{L}_2$, but not by $\mathcal{L}_3$. 

The three point couplings are rational functions of $z_1, z_2$, and have been computed in \cite{HK:2006} 
\begin{eqnarray} \label{three point}
&& C_{z_1,z_1,z_1} = \frac{1 - (6z_1 + 5z_2) + 3 (3 z_1^2 + 3 z_1 z_2 + 2 z_2^2) }{16I(z_1,z_2)} ,  \nonumber \\
&& C_{z_1,z_1,z_2} =   \frac{1 - 3z_1 - 5z_2  }{16I(z_1,z_2)} ,
\end{eqnarray}
and the others are related by symmetry.

\section{Genus one formulae} 

For the genus one case, we have the following exact  formulae from refined topological string
\begin{eqnarray} 
 \mathcal{F}^{(0, 1)} &=& -\frac{1}{2} \log(\det(\frac{\partial t_i}{\partial z_j})) -\frac{1}{12} \log(z_1z_2) - \frac{1}{4}\log I(z_1,z_2)+ \frac{1}{3}\log J(z_1,z_2) , \nonumber \\
  \mathcal{F}^{(1,0)} &=&  - \frac{1}{24} \log(z_1 z_2) - \frac{1}{12}\log J(z_1,z_2),
\end{eqnarray}
where $I, J$ are the singular divisors in (\ref{singulardivisors}). These formulae can be checked perturbatively with the expansions (\ref{perturbativeF10}), (\ref{perturbativeF01}). For the case of  $\mathcal{F}^{(0, 1)}$ amplitude, the exact formula can be also derived from the loop equation in matrix model \cite{Chekhov}. However, the case of $\mathcal{F}^{(1, 0)}$ amplitude, which comes from the $\beta$-deformation, seems much more challenging to obtain exactly from the matrix model methods. 
 
In \cite{Aganagic:2011}, the perturbative expansion for  $\mathcal{F}^{(1, 0)}$ amplitude is used to test the idea of quantum integrability in the Nekrasov-Shatashvili (NS) limit \cite{NS, MM}. Here we should follow the method in \cite{Huang:2012}, and check the formula exactly. First we define the deformed periods
 \begin{eqnarray}
  \tilde{t}_i = \oint_{A_i} \tilde{\lambda} dx, ~~~~   \tilde{\Pi}_i = \oint_{B_i} \tilde{\lambda} dx,
 \end{eqnarray}
where the deformed one-form differential $\tilde{\lambda} dx$ is determined by quantizing the curve 
\begin{eqnarray}
p^2 \Psi(x) = [W^\prime(x)^2 + f(x)]\Psi(x) ,
\end{eqnarray}
where $\Psi(x) \equiv \exp(\frac{1}{\epsilon} \int^x \tilde{\lambda} dx )$ is a wave function. The canonical quantization relation is imposed so that the operator $p=\epsilon\partial_x$.  We can use the WKB expansion of the wave function  
\begin{eqnarray}
\tilde{\lambda}(x) = \sum_{n=0}^{\infty} \lambda_n(x)  \epsilon^n,  
\end{eqnarray} 
and solve  perturbatively for the first few terms as follows
\begin{eqnarray}
&& \lambda(x) \equiv \lambda_0(x) = \sqrt{W^\prime(x)^2 + f(x)}, \nonumber \\  
&& \lambda_1(x) = -2\frac{\lambda^{\prime}(x)}{\lambda(x)}, ~~~~
 \lambda_2(x) = \frac{2\lambda^{\prime\prime}(x) \lambda(x) -3 \lambda^\prime(x)^2 }{8 \lambda(x)^3}, ~~~\cdots 
\end{eqnarray}
The odd terms are total derivatives and vanish when integrating over a contour. We are interested in the even term contributions. To compute the contour integrals, we should relate the integrand $\lambda_{2n}(x)$ to the action of some differential operators on $\lambda(x)$ plus some total derivatives of $x$. Such a differential operator is found in \cite{Aganagic:2011}, whose derivatives are with respect to the roots $x_i (i=1,2,3,4)$ of the quartic equation $W^\prime(x)^2 + f(x)=0$. Here we should write the operator as derivatives of complex structure parameters $z_1, z_2$, which is more convenient for calculations. Furthermore, we find that because of the extra Picard-Fuchs operator with non-vanishing residue, the operator is not uniquely determined but is ambiguous by addition of this operator.

We consider the differential operator without the constant and the $\partial_{z_1}\partial_{z_2}$ terms, which can be always eliminated using the Picard-Fuchs operators in (\ref{PFoperators}).  Up to a total derivative denoted as $h_2^\prime(x)$, we find $\lambda_2 = \mathcal{D}_2 \lambda + h_2^\prime(x) $, where the operator 
\begin{eqnarray} \label{D2operator}
\mathcal{D}_2 &= &  1/ [6 J(z_1,z_2)^3 (1 - 5 z_1 -5 z_2 + 
    6 z_1^2 + 4 z_1 z_2 + 6 z_2^2) ] \{ [1 + 12  (z_1 - z_2) - 
     2  (121 z_1^2 + 80 z_1 z_2  \nonumber \\ && - 25 z_2^2)   +
     4 (331 z_1^3 + 543 z_1^2 z_2 + 229 z_1 z_2^2 - 
        15 z_2^3) - 
     (3303 z_1^4 + 8084 z_1^3 z_2 + 7978 z_1^2 z_2^2 \nonumber \\ && + 
        2772 z_1 z_2^3 + 135 z_2^4)   +
     16  (243 z_1^5 + 756 z_1^4 z_2 + 1123 z_1^3 z_2^2 + 
        909 z_1^2 z_2^3 + 270 z_1 z_2^4 + 27 z_2^5) \nonumber \\ && - 
     12  (144 z_1^6 + 525 z_1^5 z_2 + 967 z_1^4 z_2^2 + 
        1314 z_1^3 z_2^3 + 894 z_1^2 z_2^4 + 225 z_1 z_2^5 + 
        27 z_2^6)]   \partial_{z_1}    \nonumber \\ && + [-1 + 2  (13 z_1 + 7 z_2) - 
  2 (100 z_1^2 + 141 z_1 z_2 + 39 z_2^2) + 
  4  (169 z_1^3 + 388 z_1^2 z_2 + 285 z_1 z_2^2   \nonumber \\ && + 
     54 z_2^3) - 
  3  (353 z_1^4 + 1098 z_1^3 z_2 + 1356 z_1^2 z_2^2 + 
     678 z_1 z_2^3 + 99 z_2^4) + 
  6  (105 z_1^5 + 403 z_1^4 z_2  \nonumber \\ && + 694 z_1^3 z_2^2 + 
     594 z_1^2 z_2^3 + 225 z_1 z_2^4 + 27 z_2^5)] \partial_{z_1}^2 \}  + (z_1\leftrightarrow z_2). 
\end{eqnarray}
There is still an ambiguity due to linear combination of the Picard-Fuchs operators $\mathcal{L}_2$ and $\mathcal{L}_3$.  We will consider the NS limit and use the perturbative expansion of $\mathcal{F}^{(1,0)}$ amplitude (\ref{perturbativeF10}), in order to fix the ambiguity. The operator $\mathcal{D}_2 $ turns out to be the correct operator which computes the deformed period from the leading term. So we find 
\begin{eqnarray}
\tilde{t}_i &=&  t_i  +(\mathcal{D}_2 t_i) \epsilon^2  +\mathcal{O}(\epsilon^4), \nonumber \\ 
\tilde{\Pi}_i &=&  \Pi_i  +(\mathcal{D}_2 \Pi_i) \epsilon^2 +\mathcal{O}(\epsilon^4). 
\end{eqnarray}

We denote the deformed prepotential in the NS limit as 
\begin{eqnarray}
\tilde{\mathcal{F}}^{(0)} ( \tilde{t}_i )= \sum_{n=0}^\infty \mathcal{F}^{(n,0)}(t_i)  (\frac{\epsilon}{4})^n,
\end{eqnarray}
 where we use a normalization factor of $4^n$ for later convenience.  As in \cite{Huang:2012} we expand the equation for $\tilde{\mathcal{F}}^{(0)}$ in terms of deformed periods, to derive differential equations for higher genus terms in the NS limit
 \begin{eqnarray}
0 &=& \partial_{\tilde{t}_i} \tilde{\mathcal{F}}^{(0)} ( \tilde{t}_i ) - \tilde{\Pi}_i  \nonumber \\
&=&  \partial_{t_i} \mathcal{F}^{(0,0)} - \Pi_i +\epsilon^2 [\frac{\partial_{t_i} \mathcal{F}^{(1,0)}}{4} 
+ (\mathcal{D}_2 t_j)( \partial_{t_i} \partial_{t_j} \mathcal{F}^{(0,0)}) -  \mathcal{D}_2 \Pi_i ] +\mathcal{O}(\epsilon^4).
 \end{eqnarray} 
After some algebra, the $\epsilon^2$ order equation can be shown to be equivalent to 
\begin{eqnarray} \label{NSF10}
\frac{1}{4} \partial_{z_i} \mathcal{F}^{(1,0)}  = C_{z_i z_j z_k} p_{z_j z_k} ,
\end{eqnarray}
where $C_{z_i z_j z_k}$ are the three-point Yukawa couplings, and $p_{z_j z_k}$ are the coefficients of $\partial_{z_j}\partial_{z_k}$ in  the differential operator $\mathcal{D}_2$.  We check the above equation (\ref{NSF10}) is satisfied exactly, using the three point functions (\ref{three point}) and the differential operator $\mathcal{D}_2$ in (\ref{D2operator}).

\section{Higher genus} 
The refined holomorphic anomaly equation \cite{Krefl:2010, Huang:2010} for higher genus $n+g\geq 2$ is a simple generalization of the original BCOV holomorphic anomaly equation 
\begin{eqnarray} \label{generalizedBCOV00}
\bar{\partial}_{\bar{z_i}} \mathcal{F}^{(n,g)}= \frac{1}{2}\bar{C}_{\bar{z_i}}^{z_jz_k}\big{(}D_{z_j}D_{z_k} \mathcal{F}^{(n,g-1)}
+{\sum_{r_1,r_2} }^{\prime}  D_{z_j} \mathcal{F}^{(r_1,r_2)}D_{z_k} \mathcal{F}^{(n-r_1,g-r_2)}\big{)}
\end{eqnarray}
where the prime denotes that the sum over $r_1,r_2$ does not include $(r_1,r_2)=0$ and $(r_1,r_2)=(n,g)$, and  the amplitudes e.g. $\mathcal{F}^{(r_1,r_2)}$ on the right hand side are understood to be zero when the indices $r_1<0$ or $r_2<0$. For the local model that we consider here, there is no contribution to the covariant derivative from the Kahler potential in the form $\partial_{z_i}K$. So the covariant derivatives in the second term in the r.h.s. of (\ref{generalizedBCOV00}) are the same as ordinary derivatives. For the first term, the covariant derivative is $D_{z_j}D_{z_k} \mathcal{F} = \partial_{z_j}\partial_{z_k} \mathcal{F} -\Gamma^{z_i}_{z_j z_k} \partial_{z_i} \mathcal{F}$, where the Christoffel connection is defined by the Kahler metric as $\Gamma_{z_i z_j}^{z_k}  =  G^{z_k \bar{z_l}} \partial_{z_i} G_{z_j \bar{z_l}}$. 

The refined topological string amplitudes at genus $n+g\geq 2$ can be written as polynomials of the propagators $S^{z_i z_j}$ of degrees $2n+3g-3$, with rational functions of complex structure moduli $z_{1,2}$ as coefficients \cite{AL, YY}. Here for the local geometry we can set the Kahler potential to be constant and only need the double-index $S^{z_i z_j}$ propagators. The propagators are defined in \cite{BCOV} by their anti-holomorphic derivative relation with the three point functions $ \partial_{\bar{z_k}}  S^{z_i z_j} =\bar{C}_{\bar{z_k}}^{z_iz_j}$ .  Using the well known special geometry relation, we can integrate  
\begin{eqnarray} \label{0624.5.2}
S^{z_i z_j} C_{z_jz_k z_l}= - \Gamma_{z_k z_l}^{z_i} + f_{z_k z_l}^{z_i}, 
\end{eqnarray}
where the $f_{z_k z_l}^{z_i}$ are holomorphic ambiguities from the integration, rational functions of $z_{1,2}$, and have been determined in \cite{HK:2006} as follows
\begin{eqnarray} \label{ambif}
f^{z_1}_{z_1z_1}&=&-\Big[6-(49z_1+48z_2)+(163z_1^2+219z_1z_2+126z_2^2) \nonumber
\\  && +(210z_1^3+304z_1^2z_2+242z_1z_2^2+108z_2^3)\Big]
/(20z_1  I J),   \nonumber \\
f^{z_1}_{z_1z_2} = f^{z_1}_{z_2z_1} &=&-\Big[29-(79z_1+157z_2)+(10z_1^2+260z_1z_2+210z_2^2)\Big]/(20 I J),  \nonumber
\\
f^{z_1}_{z_2z_2} &=&\Big[7-(55z_1+68z_2)+(142z_1^2+315z_1z_2+219z_2^2)  \nonumber \\
&&-(120z_1^3+338z_1^2z_2+492z_1z_2^2+234z_2^3) \Big]/(20 z_2 I J), 
\end{eqnarray}
where the divisors $I,J$ are available in (\ref{singulardivisors}). The other $f^{z_i}_{z_jz_k}$'s  follows from the exchange symmetry  $(z_1\leftrightarrow z_2)$.  For example, we have $f^{z_2}_{z_2z_2}=f^{z_1}_{z_1z_1}(z_1\leftrightarrow z_2)$.

The holomorphic derivatives of the propagators form a closed algebra
\begin{eqnarray} \label{0624.5.3}
\partial_{z_k} S^{z_i z_j}  = C_{z_k z_m z_n}  S^{z_i z_m}  S^{z_j z_n}  
- f_{z_k z_m}^{z_i} S^{z_j z_m} -  f_{z_k z_m}^{z_j} S^{z_i z_m} +  h_{z_k}^{z_i z_j} ,
\end{eqnarray}
where $h_{z_k}^{z_i z_j} $ are also rational functions of $z_{1,2}$. We also fix them and find it is sufficient to use $J(z_1,z_2)^3$ as the denominator. 

Assuming the algebraic independence of the anti-holomorphic derivatives of the propagators,  we can write the refined holomorphic anomaly equation (\ref{generalizedBCOV00}) in terms of partial derivatives with respect to propagators as
\begin{eqnarray} \label{holo5.4}
\partial_{S^{z_iz_j}}\mathcal{F}^{(n,g)}= \frac{1}{2} \big{(}D_{z_i}\partial_{z_j} \mathcal{F}^{(n,g-1)}
+{\sum_{r_1,r_2} }^{\prime}  \partial_{z_i} \mathcal{F}^{(r_1,r_2)}\partial_{z_j} \mathcal{F}^{(n-r_1,g-r_2)}\big{)}. 
\end{eqnarray} 
We note that because of the symmetry of propagators $S^{z_i,z_j}=S^{z_j,z_i}$, we can use only the propagators $S^{z_i,z_j}$ with $i\leq j$ in the polynomial ansatz for $\mathcal{F}^{(n,g)}$. As a result, the partial derivative $\partial_{S^{z_iz_j}}\mathcal{F}^{(n,g)}$ in the equation (\ref{holo5.4}) should be multiplied by a factor of $\frac{1}{2}$  for the case of $i\neq j$ due to double counting.  

The derivatives of the genus one amplitudes can be written as 
\begin{eqnarray} \label{derigenusone}
\partial_{z_i} \mathcal{F}^{(0,1)} &=& \frac{1}{2} S^{z_jz_k} C_{z_iz_jz_k} +\frac{1}{15} \partial_{z_i} \log(z_1z_2J^2),
\nonumber \\
\partial_{z_i} \mathcal{F}^{(1,0)} &=& -\frac{1}{24} \partial_{z_i} \log(z_1z_2J^2). 
\end{eqnarray}
We see that the derivative of the amplitude $\mathcal{F}^{(0,1)}$ is a linear function of the propagators with rational function coefficients, while the derivative of the amplitude $\mathcal{F}^{(1,0)}$ is simply a rational function of $z_{1,2}$. Utilizing the equations (\ref{0624.5.2}, \ref{0624.5.3}), the right hand side of (\ref{holo5.4}) can be computed as a polynomial of the propagators with rational functions of complex structure moduli $z_{1,2}$ as coefficients. So we can integrate this  equation and fix the coefficients of the propagators in the refined amplitude $\mathcal{F}^{(n,g)}$. For example, in the simplest case of the amplitude $\mathcal{F}^{(2,0)}$, the holomorphic anomaly equation (\ref{holo5.4}) is 
\begin{eqnarray}
\partial_{S^{z_iz_j}}\mathcal{F}^{(2,0)} = \frac{1}{2} (\partial_{z_i} \mathcal{F}^{(1,0)} ) (\partial_{z_j} \mathcal{F}^{(1,0)} ),
\end{eqnarray}
where r.h.s. is simply a rational function, found in (\ref{derigenusone}). We can easily integrate the equation 
\begin{eqnarray} \label{F20}
\mathcal{F}^{(2,0)} &=& \frac{1}{2} S^{z_iz_j} (\partial_{z_i} \mathcal{F}^{(1,0)}) (\partial_{z_j} \mathcal{F}^{(1,0)})+ f^{(2,0)}(z_1,z_2), 
\end{eqnarray} 
where  the integration constant  $f^{(2,0)}$, known as the holomorphic ambiguity, is the remaining piece that is not fixed by the holomorphic anomaly equation.

We can further fix the holomorphic ambiguity by the gap boundary conditions near the singular divisors of the Dijkgraaf-Vafa geometry. Here the ansatz for holomorphic ambiguity at genus $(n,g)$ is 
\begin{eqnarray} \label{ansatz} 
f^{(n,g)} = \frac{h^{(n,g)}(z_1,z_2)}{ (z_1z_2 J^2)^{2n+2g-2}}, 
\end{eqnarray} 
where $h^{(n,g)}(z_1,z_2)$ is a symmetric polynomial of  $z_1$ and $z_2$.  The regularity condition of  the refined amplitude around $z_{1,2}\sim \infty$ implies that the degree of  $h^{(n,g)}(z_1,z_2)$ is no higher than $12(n+g-1)$. However the situation turns out to be a little better. In \cite{Klemm:2010}, it is found empirically that the degree of the polynomial $h^{(0,n+g)}(z_1,z_2)$ in the unrefined case is no higher than $9(n+g-1)$. The refined amplitude $\mathcal{F}^{(n,g)}$ at genus $(n,g)$ has similar boundary behavior as the unrefined amplitude  $\mathcal{F}^{(0,n+g)}$, so we should expect that $h^{(n,g)}(z_1,z_2)$ has degree no higher than $9(n+g-1)$. It turns out that this is indeed the case, i.e. we can fix the holomorphic ambiguity with the gap boundary conditions, using the (\ref{ansatz}) with $h^{(n,g)}(z_1,z_2)$ a symmetric polynomial of degree $9(n+g-1)$. Although there is no rigorous proof, this empirical fact significantly simplifies the calculations,  so we can use it at low genus with precaution as long as it works.

To utilize the gap condition near the conifold divisor $J$, we choose a point on the divisor from its intersection with the line $z_1=z_2$. Near the point $(z_1,z_2)=(\frac{1}{8},\frac{1}{8})$, the good local coordinate is 
\begin{eqnarray}
z_{c,1} = z_1-z_2,~~~ z_{c,2} = 1-4(z_1+z_2). 
\end{eqnarray}
There are 3 power series solutions to the Picard-Fuchs equation near this point
\begin{eqnarray}
w_1 &=& z_{c,1} \sqrt{1+z_{c,2}}, \nonumber \\
w_2 &=& z_{c,2}^2 + 4z_{c,1}^2 z_{c,2} + (4z_{c,1}^4 +3z_{c,1}^2 z_{c,2}^2 + \frac{3}{32} z_{c,2}^4) +\mathcal{O}(z_c^ 5)  , \nonumber \\
w_3 &=& (4 z_{c,1}^3 + 3 z_{c,1} z_{c,2}^2)  + (18 z_{c,1}^3 z_{c,2} + \frac{3}{2} z_{c,1} z_{c,2}^3) +\mathcal{O}(z_c^ 5) . 
\end{eqnarray} 
One chooses the first two solutions as the flat coordinates $t_{c,i} = w_i$ near this point  \cite{Klemm:2010}.

The singular terms of the refined topological string amplitudes expanded near the divisors $z_1z_2=0$ and $J=0$ are 
\begin{eqnarray} \label{gapcondition}
\mathcal{F}^{(n,g)} &=& c^{n,g} (\frac{1}{t_1^{2n+2g-2}} +\frac{1}{ t_2^{2n+2g-2}})+ \mathcal{O}(t_1^0, t_2^0) ,  \nonumber \\ 
\mathcal{F}^{(n,g)} &=& \frac{(-2)^{11(n+g-1)}  c^{n,g}}{t_{c,2}^{2n+2g-2}}  + \mathcal{O}(t_{c,1}^0, t_{c,2}^0) . 
\end{eqnarray}
Here the factor of $(-2)^{11(n+g-1)}$ is due to a normalization of the flat coordinate $t_{c,2}$, and the constants $c^{n,g}$ appear e.g. in (\ref{perturbativeF02}, \ref{perturbativeF11}, \ref{perturbativeF20}) for genus two. These gap conditions (\ref{gapcondition}) are sufficient to fix the holomorphic ambiguities, since any non-zero ansatz in (\ref{ansatz}) with
$h^{(n,g)}(z_1,z_2) $ a polynomial of degree less than $12(n+g-1)$ can not cancel all the poles in the denominators, so would affect the singular terms.  

We fix the ambiguities and find the exact refined formulae for the genus 2 amplitudes. For example, the holomorphic ambiguity in the formula (\ref{F20}) for $\mathcal{F}^{(2,0)}$ is 
\begin{eqnarray}
 f^{(2,0)} &=& \frac{1}{720 z_1^2z_2^2 J^4}[ 
 - (11 z_1^2 + 7 z_1 z_2 + 11 z_2^2) + (231 z_1^3 +   389 z_1^2 z_2 + \cdots )  - (2079 z_1^4 + 5547 z_1^3 z_2 \nonumber \\ &&
 +  4604 z_1^2 z_2^2 + \cdots)  +  15 (693 z_1^5 + 2487 z_1^4 z_2 + 2324 z_1^3 z_2^2 + \cdots)  
    - 9 (3465 z_1^6 + 15405 z_1^5 z_2  \nonumber \\ && + 19435 z_1^4 z_2^2 + 
    8382 z_1^3 z_2^3 +\cdots)  + 
 3 (18711 z_1^7 + 98037 z_1^6 z_2 + 172791 z_1^5 z_2^2 + 
    21757 z_1^4 z_2^3 \nonumber \\ &&  +\cdots )   -  3 (18711 z_1^8 + 111699 z_1^7 z_2 + 266274 z_1^6 z_2^2 + 
    50973 z_1^5 z_2^3 - 272722 z_1^4 z_2^4  +\cdots)  \nonumber \\ && 
    +  27 (z_1 - z_2)^2 (891 z_1^7 + 7695 z_1^6 z_2 + 
    32625 z_1^5 z_2^2 + 67333 z_1^4 z_2^3+\cdots )], 
\end{eqnarray}
where the $\cdots$ denote the terms implied by the symmetry $z_1 \leftrightarrow z_2$. The formulas for $\mathcal{F}^{(0,2)}$ and $\mathcal{F}^{(1,1)}$ are listed in Appendix \ref{formulas.6.24}. 

We expand the exact formula in terms of the flat coordinates $t_{1,2}$, using the formula for propagators (\ref{0624.5.2}) and inverting the expansions (\ref{periods.6.24}). Here the Christoffel connections can be computed in the holomorphic limit by $\Gamma_{z_i z_j}^{z_k}  =   (\partial_{t_l}  z_k)( \partial_{z_i}  \partial_{z_j}  t_l)$. We checked the expansions near the point $(z_1,z_2)=(0,0)$ agree with the higher order terms from perturbative matrix model calculations in  (\ref{perturbativeF02}, \ref{perturbativeF11}, \ref{perturbativeF20}). Our exact formulas provide a much more efficient way to compute the higher order terms than the perturbative method in matrix model.

\vspace{0.2in} {\leftline {\bf Acknowledgments}}

We thank Albrecht Klemm for discussions and collaborations on related papers. MH is supported by the ``Young Thousand People" plan by the Central Organization Department in China.

\appendix 
\section{The formulas for $\mathcal{F}^{(0,2)}$ and $\mathcal{F}^{(1,1)}$} 
\label{formulas.6.24}
In the Appendix we write down the formulas for the other genus two amplitudes. The unrefined formula $\mathcal{F}^{(0,2)}$ has been obtained in the previous paper \cite{HK:2006}. Here we rewrite it in the polynomial formalism. To make the expression compact, we do not write the coefficients of the propagator polynomials completely as the  explicit rational functions of $z_{1,2}$. Instead, we use various ingredients such as the three-point Yukawa couplings $C_{ijk}$ in (\ref{three point}), the derivatives of the amplitude $\mathcal{F}^{(1,0)}$ in (\ref{derigenusone}) and the rational functions $f^i_{jk}$, $h^{ij}_k$  appeared in (\ref{ambif}, \ref{0624.5.3}).

The amplitudes $\mathcal{F}^{(1,1)}$ and $\mathcal{F}^{(0,2)}$ are respectively quadratic and cubic polynomials of the propagators, from the integration of the holomorphic anomaly equation. The expressions are  
\begin{eqnarray}
\mathcal{F}^{(1,1)} &=& \frac{1}{2} S^{ij}S^{kl}C_{ijk}\partial_l  \mathcal{F}^{(1,0)} + \frac{1}{2}S^{ij}( \partial_i\partial_j \mathcal{F}^{(1,0)} -f^k_{ij} \partial_k \mathcal{F}^{(1,0)})   \nonumber \\
&& +\frac{1}{15}S^{ij} (\partial_i \mathcal{F}^{(1,0)}) 
\partial_j \log(z_1z_2J^2) +f^{(1,1)} (z_1,z_2),  \\
\mathcal{F}^{(0,2)} &=& S^{ij}S^{kl}S^{mn} ( \frac{1}{12} C_{ikm}C_{jln} + \frac{1}{8} C_{ijk}C_{lmn}) +\frac{1}{30}S^{ij}S^{kl} C_{ijk}  (\partial_l \mathcal{F}^{(1,0)})  \nonumber \\
&& +\frac{1}{16} S^{ij}S^{kl} (\partial_i C_{jkl}+ \partial_j C_{ikl} -2 f^m_{ij}C_{mkl}) -\frac{1}{4}S^{ij}S^{kl} f_{ik}^mC_{mjl} \nonumber \\ && 
+S^{ij}[ \frac{1}{4}  C_{ikl}h^{kl}_j+ \frac{1}{30}\partial_i\partial_j \log(z_1z_2J^2)  -\frac{1}{30}  f^k_{ij} \partial_k \log(z_1z_2J^2) ] \nonumber \\ && +\frac{S^{ij} }{450} \partial_i \log(z_1z_2J^2) \partial_j \log(z_1z_2J^2)  
+ f^{(0,2)}(z_1,z_2) . 
\end{eqnarray} 

The holomorphic ambiguities are fixed by the gap boundary conditions to be the followings 
\begin{eqnarray}
f^{(1,1)} &=& \frac{1}{900z_1^2z_2^2J^4} [(8 z_1^2 + 28 z_1 z_2 + 8 z_2^2) - 
 7 (24 z_1^3 + 137 z_1^2 z_2 + \cdots)  + 
 2 (756 z_1^4 + 5937 z_1^3 z_2 \nonumber \\ &&  + 7450 z_1^2 z_2^2 +  \cdots)  
 - 45 (168 z_1^5 + 1669 z_1^4 z_2 + 2403 z_1^3 z_2^2 +  \cdots)  + 
 24 (945 z_1^6 + 11295 z_1^5 z_2  \nonumber \\ &&  + 18585 z_1^4 z_2^2  + 
    15998 z_1^3 z_2^3  + \cdots)  - 
 27 (1512 z_1^7 + 21027 z_1^6 z_2 + 39767 z_1^5 z_2^2 + 
    23710 z_1^4 z_2^3   \nonumber \\ && + \cdots ) + 
 162 (252 z_1^8 + 3981 z_1^7 z_2  + 8646 z_1^6 z_2^2 + 
    3379 z_1^5 z_2^3 - 3844 z_1^4 z_2^4 +  \cdots)  
\nonumber  \\  &&    - 81 (z_1 - z_2)^2 (216 z_1^7 + 4239 z_1^6 z_2 + 
      17667 z_1^5 z_2^2  + 34198 z_1^4 z_2^3 +\cdots)   ] , \\
f^{(0,2)} &=& \frac{1}{9000z_1z_2J^2} [ -1253 + 10503  (z_1 + z_2) -  27 (1081 z_1^2 + 950 z_1 z_2 + 1081 z_2^2)
  \nonumber  \\&& +  26865  (z_1 + z_2) (z_1 - z_2)^2 ] ,
\end{eqnarray}
where $\cdots$'s denote terms implied by the exchange symmetry $z_1\leftrightarrow z_2$.

\end{document}